\documentclass[pre,twocolumn,showpacs,superscriptaddress] {revtex4}
\usepackage{graphics}
\newcommand{\cu}
{\affiliation{Department of Physics, University of Calcutta,
92 Acharya Prafulla Chandra Road, Kolkata 700009, India}}

\begin{document}
  
 \title
 { A new model of binary opinion dynamcis:  coarsening and effect of 
disorder}
 \author
 {
Soham Biswas}
\cu
\author	{  Parongama Sen }
\cu
\begin{abstract}

We propose a model of binary opinion 
in which the opinion of the individuals change according to 
 the state of their neighbouring 
domains.
If the neighbouring domains have opposite opinions, then the opinion of the 
domain with 
the larger size is followed.
Starting from a random configuration, the system evolves to   a homogeneous 
state. The dynamical evolution  show novel scaling behaviour with the 
 persistence exponent $\theta \simeq 0.235$
and dynamic exponent $z \simeq1.02 \pm 0.02$.
Introducing disorder through a parameter called rigidity coefficient $\rho$ 
(probability that people are completely rigid and never change 
their opinion), 
the transition to a heterogeneous society  
at $\rho = 0^{+}$ is obtained.
Close to $\rho =0$, the   equilibrium values of the 
dynamic variables show power law scaling behaviour with $\rho$.
 We also discuss the effect of having both quenched and annealed disorder in the system.


\vskip 0.5cm

\end{abstract}

\maketitle

A number of models which simulate the formulation
of   opinion in a social system have been proposed in the physics
literature recently \cite{Stauff}.  Many of these show a  close connection to familiar models of
statistical physics, e.g.,
the  Ising and the Potts models.
Different kinds of phase transitions have also been observed
in these models by introducing suitable parameters.
One such phase transition can be from a homogeneous society
where everyone has the  same opinion to
a heterogeneous  one with mixed opinions \cite{phase-tr}.

In  a model of opinion dynamics, the key feature is
the interaction of the individuals. Usually, in all the models, it is assumed that
an indiviual
is influenced by its nearest neighbours.
In this paper we propose a one dimensional 
model of binary opinion in which the dynamics 
is dependent on the {\it {size}} of the neighbouring domains as well.
Here an individual changes his/her opinion in two situations: 
first when the two  
neighbouring domains have opposite polarity,  and in this case 
the individual simply follows the opinion of
the neighbouring domain with the  larger size.
This case may arise only when the individual is at the boundary of the two 
domains.
An individual also 
changes his/her opinion when both the  neighbouring domains have an opinion 
which opposes his/her original opinion, i.e., the individual is  
sandwiched between two domains of same polarity.
It may be noted that for  the second case, 
the size of the neighbouring domains is irrelevant.
 
\begin{figure}[ht]
{\resizebox*{6cm}{!}{\includegraphics{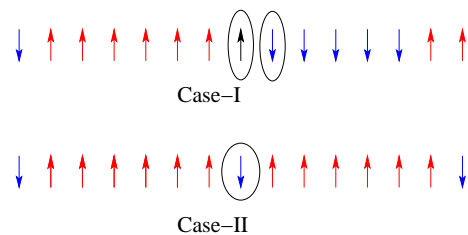}}}
\caption{(Color online) Dynamical rules for Model I: in both cases the encircled spins may change state; in case I, 
the boundary spins 
will follow the opinion of the left domain of up spins which  will grow. 
 For case II,  the 
down spin between the two up spins will flip irrespectative of the size of the neigbouring domains.}
\label{model}
\end{figure}

This model, henceforth referred to as Model I, can 
be represented by a system of Ising spins where the up and down states
correspond to the two possible opinions. 
The  two  rules followed in the dynamical evolution  
  in the equivalent spin model 
are  shown schematically 
in Fig. \ref{model} as case I and  II. 
In the first case the spins representing  individuals  at the boundary   
between two domains 
will choose the opinion of the 
left side domain (as it is  larger in size). For the 
second case the down spin  flanked by  two neighbouring  up spins 
will flip.

 The main idea in Model I is that the size of a domain represents
 a quantity analogous to `social pressure' which is expected 
  to be proportional to the  number of  people 
supporting a cause. An individual, sitting at the domain boundary, is most exposed to
the competition between  opposing pressures and  gives in to the larger one.
This is what happens in case I shown in Fig.\ref{model}. 
The interaction in case II on the other hand implies that it is difficult to 
stick to   one's opinion if the entire neighbourhood opposes it.  

Defining the dynamics in this way, one immediately notices that case II corresponds to what would
happen for  spins in a nearest neighbour ferromagnetic Ising model (FIM) in which the dynamics
at zero temperature is simply an energy minimisation scheme.
However,   the boundary spin in the FIM
behaves differently in case I; it may or may not flip as the energy remains same.
 In the present model, the dynamics is
deterministic even for the boundary spins
(barring the rare 
instance when the two neighbourhoods have the same size in which case 
the individual changes state with fifty percent probability).

In this
model, the important condition of changing one's opinion is the size of the
neighbouring domains which is not fixed  either in time or space.
This is the unique feature of this model, and to the best of our knowledge
such a condition has not been considered earlier. In the most familiar  
models of opinion dynamics like the Sznajd model \cite{Sznajd} or the voter model \cite{vote},
one takes the effect of nearest neighbours within a given radius and
even in the case of models defined on networks \cite{v-network}, 
the influencing neighbours may be
nonlocal but always fixed in identity.

We have done Monte Carlo simulations  to study the dynamical evolution of the 
proposed model from a given initial state.  With a system of $N$  spins representing individuals, 
at each step, 
one spin  is selected at random and 
its state updated. After $N$  such steps, one Monte Carlo time step is 
said to be completed.

If $N_{+}$ is the number of people of a particular opinion (up spin) 
and $N_{-}$ is the number of people of opposite opinion (down spin), 
the order parameter is defined as $m=\vert N_{+}-N_{-} \vert/N $.
 This is 
identical to the (absolute value of) magnetisation in the Ising model.

Starting from a random initial configuration, 
the  dynamics in Model I leads to a final state with $m=1$, i.e.,   
a homogeneous state where all individuals have the same opinion. 
It is not difficult to understand this result; in absence of any fluctuation, the dominating neighbourhood (domain) 
simply grows in size ultimately spanning the entire 
system.

\begin{figure}[ht]
{\resizebox*{8cm}{!}{\includegraphics{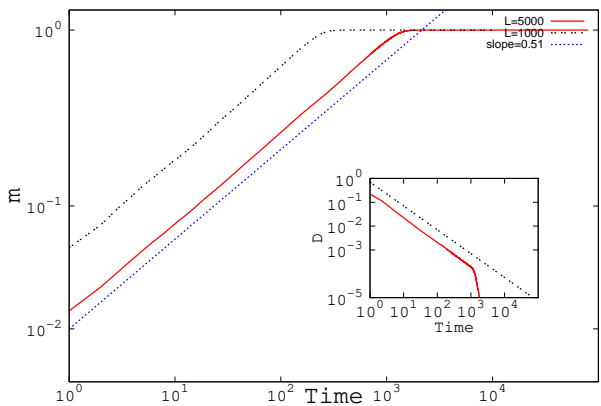}}}
\caption{ (Color online) Growth of order parameter $m$ with time for two different system sizes along with
a straight line (slope 0.51) shown in a log-log plot.
Inset shows the decay of fraction of domain wall $D$ with time.}
\label{NP_Dw}
\end{figure}

%
We have studied the dynamical behaviour of the fraction of
domain walls $D$ and  the order parameter $m$ as the 
system evolves to the homogeneous state.
We observe that the behaviour of $D(t)$  and $m(t)$
are consistent with the usual  scaling
behaviour found in
coarsening phenomena;
  $D(t) \propto  t^{-1/z}$ with $z = 1.00 \pm 0.01 $ and  $m(t) \propto t^{1/2z}$
with  $z = 0.99 \pm 0.01 $.
These variations are shown in Fig. \ref{NP_Dw}.

We have also calculated the persistence probability 
that a person has not 
change his/her opinion up to time $t$.
Persistence, which in general is the  probability       
that a fluctuating nonequilibrium field does not change sign upto time $t$, 
shows a power law decay behaviour 
in many physical phenomena, 
i.e., 
$P(t) \propto t^{-\theta}$, 
where $\theta$ is the  persistence exponent.
In that case, one can use the finite 
size scaling relation \cite{puru,biswas_sen}
\begin{equation}
P(t,L) \propto t^{- \theta}f(L/t^{1/z}).
\label{fss}
\end{equation}
For finite systems, the 
persistence probability saturates at a value $ \propto L^{-\alpha}$ at large times. 
Therefore,  for
 $x <<1$ , $f(x)  \propto x^{-\alpha}$ with $\alpha = z\theta$.  For large $x$,
$f(x)$ is a constant. Thus one can obtain  estimates for both  $z$ and $\theta$ using the above
scaling form.

\begin{figure}[ht]
{\resizebox*{8cm}{!}{\includegraphics{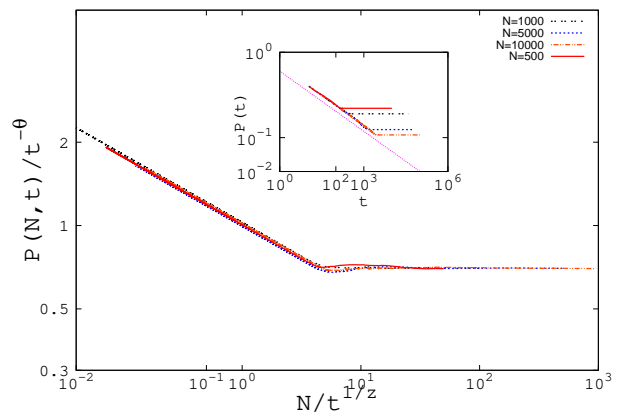}}}
\caption{(Color online) The collapse of scaled persistence probability versus scaled time using $\theta=0.235$ and $z=1.04$ is shown for
different system sizes.  Inset shows 
the unscaled data.}
\label{per_noparam}
\end{figure}

In the present model  the  persistence 
probability does show a power law decay with $\theta =0.235\pm0.003$, 
while the finite size scaling analysis made according to (\ref{fss}) 
suggests a $z$ value $1.04 \pm 0.01$ (Fig.  \ref{per_noparam}).
Thus we find that the values of $z$ from the three different 
calculations are consistent and conclude that the dynamic exponent 
$z = 1.02 \pm 0.02$. 



It is important to note that both the exponents $z$ and $\theta$ 
are novel in the sense 
that they are   quite different from those of  
the one dimensional Ising model \cite{Derrida} and other 
opinion/voter dynamics models \cite{stauffer2,sanchez,shukla}. 
Specifically in the Ising model,  
$z=2$ and $\theta = 0.375$   
and for the   Sznajd model
the persistence exponent is  equal to that of the Ising model.
  This 
shows that the present model belongs to an entirely  new dynamical class.

The Model I described so far    has no fluctuation. 
Fluctuations or disorder can be introduced in several 
ways.
We adopt a realistic outlook:   since every individual is not expected to 
succumb to social pressure, we  
 modify  Model I by introducing a parameter $\rho$ called rigidity 
coefficient which denotes the probability that people are completely rigid and never change their opinions.
The modified model will be called Model II in which 
there are  $\rho N$ rigid individuals  (chosen randomly at time $t=0$),  who  
retain their initial state 
throughout the time evolution. 
Thus the disorder is quenched in nature. 
The limit $\rho=1$
corresponds to the unrealistic  noninteracting case when no time evolution 
takes place; 
$\rho=1$ is in fact a trivial fixed point.
For other values of $\rho$, the system evolves to a equilibrium state.

\begin{figure}[ht]
{\resizebox*{8cm}{!}{\includegraphics{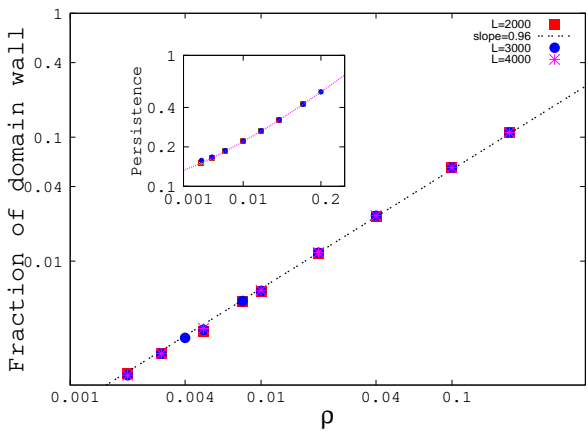}}}
\caption{(Color online) Satuaration values   of fraction of domain walls  $D_s$  and  persistence probability $P_s$ (shown in inset) 
increase with rigidity coefficient $\rho$ in a power law manner. There is no system size dependence for both the quantities.}
\label{sat_domain_per}
\end{figure}

\begin{figure}[ht]
{\resizebox*{8cm}{!}{\includegraphics{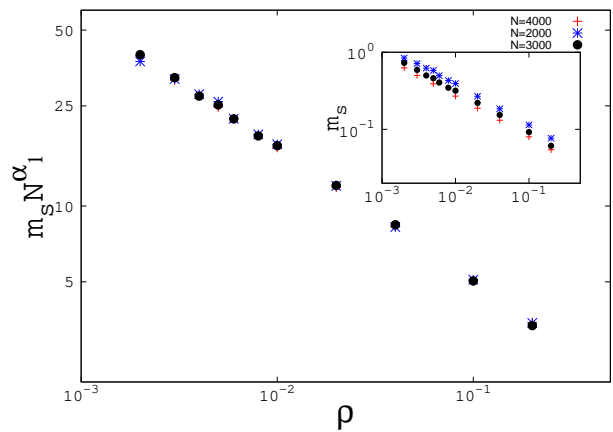}}}
\caption{(Color online) Scaled satuaration value of $m_s$  decays with the rigidity coefficient $\rho$. 
 nset shows the unscaled data.
}
\label{sat_mag}
\end{figure}

The time evolution changes drastically in nature with the 
introduction of $\rho$. All the dynamical variables like
order parameter,  fraction of domain wall and persistence
attain a saturation value at a rate which increases with 
$\rho$.  Power law variation with time can only be observed for $\rho < 0.01$ 
with the exponent values same as those for $\rho = 0$.
The saturation or equilibrium values on the other hand show
the following behaviour:
\begin{eqnarray}
&&m_s\propto N^{-\alpha_1}\rho^{-\beta_1}\nonumber\\
&&D_s \propto \rho^{\beta_2}\nonumber\\
&&P_s  =  a + b\rho^{\beta_3}  \label{}
\end{eqnarray}
where in the last equation  $a (\sim 10^{-2})$ and $\beta_3 (\sim 0.36)$ are dependent on $N$.
The values of the exponents are $\alpha_1 = 0.500 \pm 0.002,$ $\beta_1 = 0.513 \pm 0.010$, $\beta_2= 0.96 \pm 0.01$.
(Figs \ref{sat_domain_per} and \ref{sat_mag}.)
The variation of $m_s$ with $\rho$ is strictly speaking not valid for extremely
small values of $\rho$. However, at such small values of $\rho$, it is 
difficult to obtain the exact form of the variation  numerically. 

It can be naively assumed that the $N\rho$ rigid individuals will dominantly appear at the 
domain boundaries such that  in the first order approximation (for a fixed 
population),
 $D \propto 1/\rho$. This would give $m \propto 1/\sqrt{\rho}$ 
indicating  $\beta_1 = 0.5$ and $\beta_2 =1$.   
The numerically obtained values are in fact quite close
to these estimates.

The results obtained for Model II can be explained in the following way:
with $\rho \neq 0$, 
the domains cannot grow freely  and domains with both
kinds of opinions survive making the equilibrium 
$m_s$ less than unity.
Thus  the society becomes heterogeneous for any $\rho > 0$ when people
do not follow the same opinion any longer. 
The variation of $m_s$ with $N$ shows  
that $m_s \rightarrow 0$ in the thermodynamic limit for $\rho > 0$. 
 Thus not only does the society become  heterogenous at  the onset
of $\rho$, it goes to a completely disordered state 
analogous to the
paramagnetic state in magnetic systems.  
Thus one may conclude that a phase transition from a ordered state with $m=1$
to a disrodered state ($m=0$) takes place for $\rho = 0^+$. It may be
recalled  here that  $m=0$ at the trivial fixed point $\rho=1$ 
and therefore the system flows to the $\rho=1$ fixed point for any 
nonzero value of $\rho$ indicating  that $ \rho=1$ is a stable fixed point.

That the saturation values of the fraction of domain walls do not show system size dependence for 
$ \rho=0^{+}$ further
 supports the fact  that  the phase transition occurs at $\rho=0$.

%
The effect of the parameter $\rho$ is therefore very similar to thermal
fluctuations in the Ising chain, which drives the latter 
 to a disordered state for any non-zero temperature, $\rho=1$ being
comparable to infinite temperature.
However, the role of the rigid individuals
 is  more similar to 
domain walls which are pinned rather than thermal fluctuations.
In fact, the Ising model will have dynamical evolution even at very high temperatures
while in Model II, the dynamical evolution becomes slower with $\rho$, ultimately stopping altogether 
at $\rho=1$. This is reflected in the scaling of the various thermodynamical quantities with $\rho$,
e.g., the order parameter 
shows  a power law scaling 
above the transition point. 

\begin{figure}[ht]
{\resizebox*{7cm}{!}{\includegraphics{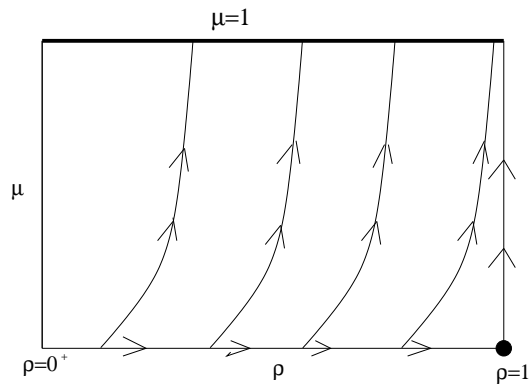}}}
\caption{
The flow lines in the $\rho-\mu$ plane:  
Any non-zero value of $\rho$ with $\mu=0$ drives
the system  to the disordered fixed point $\rho=1$. 
Any nonzero value of $\mu$ drives it to the 
ordered state ($\mu=1$, which is a line of fixed points) 
for all values of $\rho$.}
\label{flow}
\end{figure}

Since the role of $\rho$ is similar to domain wall pinning, one can
introduce a depinning probability factor $\mu$ which in this system  
represents the probability for  rigid individuals  to become non-rigid
during each Monte Carlo step. $\mu$  relaxes the rigidity
criterion in an annealed manner in the sense 
that the identity of the individuals who become non-rigid is not fixed (in time). If  
$\mu=1$, one gets back  Model I (identical to  Model II with $\rho$=0) whatever
be the value of $\rho$, and therefore $\mu=1$ signifies a line of (Model I) 
fixed points,
where the dynamics leads the system to a homogeneous state.

With the introduction of $\mu$, one has effectively  a lesser fraction  $\rho^\prime$  of
rigid people in the society,  where
\begin{equation}
\rho^\prime = \rho(1-\mu).
\end{equation}
The difference  from Model II is, of course, that 
this effective fraction of   rigid individuals 
is not fixed in identity (over time). Thus when $\rho \neq 0, \mu \neq 0$, we have a system
in which there are both quenched and annealed disorder.
It is observed that for any  nonzero value of $\mu$, 
the system once again evolves to a homogeneous state ($m=1$) for 
all values of $\rho$. Moreover, the dynamic behaviour is also same as Model I 
with the exponent $z$ and $\theta$  having identical 
values.
This shows that the nature of randomness is crucial as 
 one cannot simply 
replace a system with parameters \{$\rho \neq 0, \mu \neq 0$\} 
by one with only quenched randomness
 \{$\rho^\prime \neq 0$, $\mu^\prime=0$\} as in the latter   case
one would end up with a heterogeneous society.
We therefore conclude that the annealed disorder
wins over the quenched disorder;
  $\mu$ effectively drives the system to
 the  $\mu=1$ fixed point  for 
any value of $\rho$. This is shown   schematically in  a flow diagram (Fig. \ref{flow}).
It is worth remarking that it looks very similar to the flow
diagram of the one dimensional Ising model with nearest neighbour 
interactions in a longitudinal field and finite temperature. 

In summary, we have proposed a model of opinion dynamics in which
the social pressure is quantified in terms of the size of domains 
having  same opinion. In the simplest form,  the model has no disorder and
self organises to a homogeous state in which the entire population has
the same opinion. This simple model exhibits novel coarsening exponents.
With disorder, the model undergoes a phase transition 
from a homogeneous society (with order parameter equal to one) 
to a heterogeneous one which is fully disordered
in the sense that no consensus can be reached as the order parameter
goes to zero in the thermodynamic limit.
 With both quenched and annealed 
randomness present in the system, the annealed randomness is observed
to drive the system to a homogeneous state for any amount of the quenched
randomness.

Many open questions still remain regarding  Models I and II,
the behaviour in higher dimensions being one of them.
 In fact, full understanding of the phase transition   occurring in Model II
reported here 
is   an important issue:   although   the phase transition 
has similarities with the one
dimensional Ising model, there are some distinctive features
which should be  studied in more detail.

Acknowledgments: Financial support from DST project SR/S2/CMP-56/2007
and  stimulating discussions with P. K. Mohanty and B. K. Chakrabarti are 
acknowledged.



\begin{thebibliography}{99}

\bibitem{Stauff} D. Stauffer, arXiv:{{0705.0891}} 
\bibitem{phase-tr} 
A. Baronchelli, L. Dall'Asta, A. Barrat, and V. Loreto,
Phys. Rev. E {\bf{76}}, 051102 (2007);
C. Castellano, M. Marsili and A. Vespignani,
Phys. Rev. Lett. {\bf{85}} 3536 (2000).
\bibitem{Sznajd} K. Sznajd-Weron and  J. Sznajd,   Int.J.Mod.Phys C {\bf{11}} 1157 (2000).
\bibitem{vote} T. M. Liggett   \textit{Interacting Particle Systems: Contact, Voter and Exclusion Processes} (Springer-Verlag Berlin 1999).
\bibitem{v-network} C.Castellano, D.Vilone and A.Vespignani, Europhys. Lett. 
{\bf{63}} 153 (2003).
\bibitem{puru} G. Manoj and P. Ray, Phys. Rev. E {\bf {62}} 7755 (2000); G. Manoj and P. Ray, J. Phys A {\bf{33}} 5489 (2000).
\bibitem{biswas_sen} S.Biswas, A.K.Chandra and P.Sen, Phys. Rev. E {\bf{78}}, 041119 (2008)
\bibitem{Derrida} B. Derrida in {\it{Lecture notes in Physics}} {\bf{461}} 
165 (Springer, 1995); B. Derrida, V. Hakim, V. Pasquier, Phys. Rev. Lett. {\bf
{75}} 751 (1995)
\bibitem{stauffer2} D. Stauffer and P. M. C. de Oliveira, Eur. Phys. J B 
{\bf{30}} 587 (2002.)

\bibitem{sanchez} J.R.Sanchez, arXiv:cond-mat/{{0408518v1}} 
\bibitem{shukla} P. Shukla, J. Phys. A Math. Gen. {\bf{38}} 5441 (2005).  
 \end{thebibliography}
\end{document}